# Monitoring of slopes, rock faces and masonry walls in a 19th century public park: the example of the Buttes Chaumont Park (Paris, France).

**Marc Peruzzetto,** Isabelle Halfon, Clara Levy, Florian Masson, Aurore Ramage, Gildas Noury, Ribes David
*BRGM, Orléans, France, m.peruzzetto@brgm.fr*

Daoud Benazzouz, Marina Kudla, Laurence Lejeune
*Mairie de Paris, Direction des Espaces Verts, France*

ABSTRACT: Developed on former gypsum quarries, the Buttes Chaumont Park is a 25-hectare geotechnical complex that is unique in the world. After three years of heavy work to create, in particular, an artificial cave, a lake and an island, the park opened in 1867 and has suffered gravitational hazards ever since (landslides, rockfalls and sinkholes). The BRGM has worked with the Paris City Council since 2021 to characterize the geological and geotechnical context, identify major gravitational hazards, and monitor the evolution of instabilities in slopes and rock/masonry walls. In this context, the BRGM has proposed, defined and followed a geotechnical supervision scheme including four levels of monitoring: detailed quarterly site visits since March 2023, bimonthly tacheometric surveys (operating since December 2022), monthly manual gauges measurements (since January 2024), and automatic extensometers and temperatures measurements (since March 2024). The interpretation of the data allows to confirm and/or complement the gravitational hazard mapping that had been carried out in 2022. By analyzing the correlation between displacement measurements and meteorological conditions, we could also differentiate between seasonal/daily trends mainly associated with temperature variations, and displacements associated with gravitational processes. These results help mitigate risks in the Buttes Chaumont Park in its current state, and adapt works planned in the coming years to restore and secure the park.

## 1 INTRODUCTION

The Buttes-Chaumont Park is a 25 ha green haven in the heart of Paris 19th district. It is the largest inner-city park of Paris and is visited by about 6 million people every year. Its unique scenery results from extensive works carried out between 1864 and 1867 to remodel a former gypse quarry. In addition to large lawns and tree covered hills, the park owes its reputation to an artificial lake, island (with its well-known temple of Sybil, Figure 1), cave and waterfall. It results in a unique geotechnical complex combining natural and artificial reliefs and rock faces. Although remarkable, this complex has suffered various gravitational and geotechnical hazards since its opening, including rockfalls, landslides, sinkholes and masonry aging. Several consolidation and restoration works have been carried out since 1900. In 2021, a new expertise was carried out by the BRGM (French National Geological Survey) and the IGC (General Quarry Inspection) for the Green Spaces Department of Paris city council. It allowed to characterize the geotechnical context and history of the park through extensive archive works, and the geotechnical and gravitational hazards in the sector of the island. This preliminary work led to closing access for the public to the most vulnerable areas (in particular, the island) and laid the foundation for the most extensive restoration project since the opening of the park. A competitive dialogue has been initiated in 2023 and is still ongoing, with works scheduled to start in 2026-2027. In the meantime, the monitoring of the park is of prior importance for ensuring the security of the park users and workers as well as to quantify the seasonal/long trend of gravitational processes, and in turn, help design adequate mitigation measures. In this work, we present the monitoring framework that was designed and followed by the BRGM. In Section 2, we describe the geological and geotechnical context and associated hazards in the park. The four components of the monitoring strategy are then given in Section 3. Main findings are summarized in Section 4 and discussed in Section 5.

## 2 CONTEXT

### 2.1 *Geological and geotechnical context*

The Buttes Chaumont Park history has been thoroughly investigated by the BRGM through archive and field work (Noury, 2024). It lies at the western end of the Romainville plateau which corresponds to formations from Upper Eocene (Ludian stage, from -37.8 to -33.9 million years ago) and Lower Oligocene (Stampian stage, from -33.9 to -28.1 million years ago). The highest part of the park (to the east) lies 90-95 m above sea level. The lake, with its banks at around 58 m asl, is the lowest point in the park. The terrain at the top of the park is made up of fill, followed rapidly by Brie Limestone, Romainville Green Clay (GC) and Supra-Gypsum Marls (SGM). The underlying four gypsum masses (commonly referred to as GM1 to GM4) have been mined in multiple locations in the area since at least the 17th century, causing the hill to recede by around 150 m.

Initially carried out in the open air, gypsum mining gradually moved underground, with multiple caved-in entrances to reach the deepest masses (GM2 and GM3). Chambered quarries and turned pillars then extended beneath the plateau, with cavities reaching 20 m in height. Despite a royal decree in 1778 prohibiting caving operations, underground quarrying certainly continued for many years, legally and/or illegally. There is little information on the end of quarrying activities, but filling in of cavities started in 1780 by caving (collapse of the cavities on themselves) and later on by backfilling. Open-cast quarrying, however, continued for years until being limited by the available space.

The quarry irregular topography led to designing a scenery that mimicked the Romantic vision of nature of the second half of the 19th century. The works initiated in 1864 involved significant excavations (about 410,000m$^3$) and backfilling (about 520,000 m$^3$, particularly around the island to create an artificial 2.2 ha lake (Figure 1 and Figure 2). The island has been remodeled from a mass that hadn't been quarried, with the succession (from top to bottom) of a gypsum mass (GM1), clays (GC), marls (SGM) and fills (Figure 2). The gypsum is apparent on the western and southern sides of the island. The northern and southern sides are more complex: to create a vertical and higher rockface (up to 20 m), the gypsum is recut and overtopped by retaining walls supporting the marls and backfills. The whole face is then recovered by a millstone cladding (Figure 1). Finally, a fully artificial cavern and a waterfall are created near the southern shore of the lake in a recess of the quarry face.

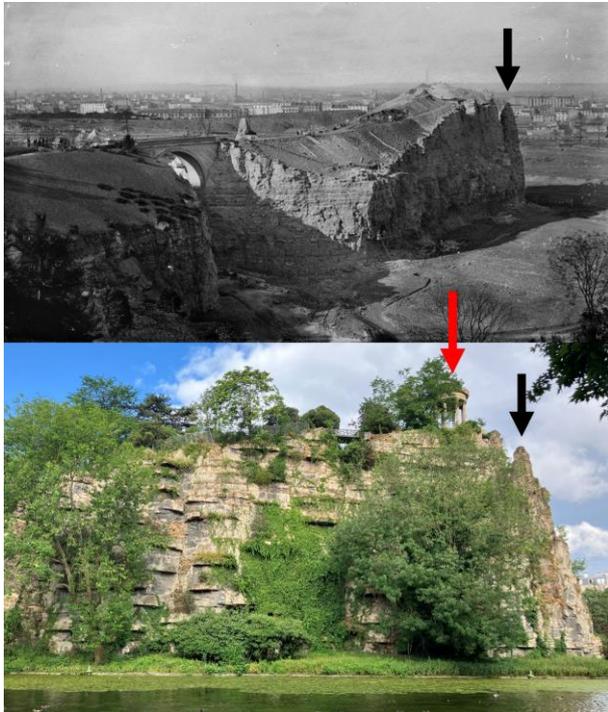

Figure 1. Picture of the eastern side of the island during construction works around 1865 (top) and in 2025 (bottom). The black arrows locate an artificial peak on both images. The red arrow locates the Temple of Sybil.

2.2  *Geotechnical and geological hazards*

Because of its unique geological and geotechnical setup, the Buttes Chaumont Park has quickly faced structural damage. The lake slab is damaged by several collapses quickly after the park opening and was progressively renovated between 1900 and 1969. The rock face of the park has also experienced several rockfalls leading to purges and consolidation works between 1886 and 1902, including masonry works on the east side of the island. A shotcrete cover is also added on the southern side of the island in the 1980s. At the same period, the artificial cavern was also significantly reinforced with concrete injections behind the structure and anchoring. Finally, several landslides occurred in the parc, especially between 1867 and 1930, affecting the upper unstable formations (backfills, green clays and marls).

There are multiple causes for these disorders:
- The earthworks carried out for the park development have probably favored the occurrence of landslides, and also the numerous collapses of the lake bottom. Concerning these collapses, the role of former underground quarries and the possible dissolution of gypsum materials should also be considered. Such events (collapses/subsidence and landslides) are often triggered by heavy water inflows (rainfall, leaks).
- The dissolution of gypsum materials and roots developments is a major driver of rockfall activity on the gypsum rock faces.
- Finally, the ageing of geotechnical structures (built 100 and 150 years ago), with chemical interactions deteriorating mortars and steel reinforcements, accentuates the probability of retaining structure failing and rockfalls.

A cartography of gravitational hazards (landslide, rockfall and cliff recessions) was carried out by the BRGM in 2022 (Halfon and Flipo-Houzé, 2022; Levy, 2022). They map:
- The areas with steep slopes and unconsolidated materials (backfills, clays and marls) as most prone to landslides;
- Diffuse rockfall hazard on rock and masonry walls, with associated possible propagation areas;
- Localized rockfall hazards associated to compartments delimited by natural cracks and identified as potentially unstable.

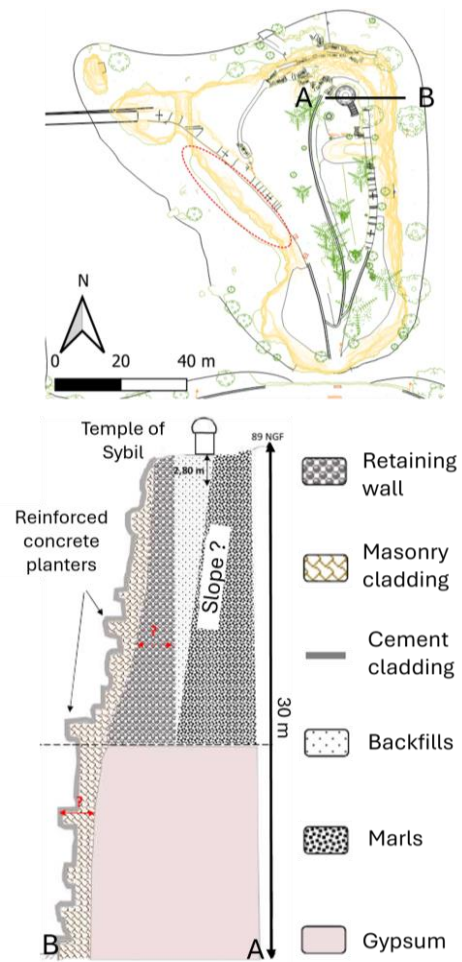

Figure 2. Top: Topographic map of the island. The red ellipse locates topographic targets from Figure 5. Bottom: Cross-section of the island eastern wall with main geological / geotechnical units (section A-B on map). The Temple of Sybil is shown in Figure 1.

3  MONITORING SCHEME

Following these first analyses, the BRGM recommended a monitoring scheme (Peruzzetto, 2023) comprising several observation levels and frequency to identify short-term and medium-term trends. This includes:
- Quarterly on-site visits (starting March 2023) by BRGM experts to identify long-term evolutions of landslides and catalogue new disorders (rockfalls, fissures, …). Sites of interest are photographed at every visit to document visible changes;
- Bimonthly tacheometric survey of about 60 targets installed on the island walls and the former quarry rockface

(operating since December 2022, carried out by an external surveyor).
- Monthly manual gauges measurements at about 20 easily accessible fissures on pathways and masonry (operating since January 2024, carried out by an engineering company);
- Automatic, continuous (every 10 min) extensometers measurements on 12 fissures on the island walls, installed by rope access technicians (operating since March 2024, maintained by an external engineering office). In this work, we focus on one vertical fissure in the gypsum mass on the western side of the island (*fissure 1*, Figure 3a), and one horizontal fissure in a retaining wall (*fissure 2*, Figure 3b).

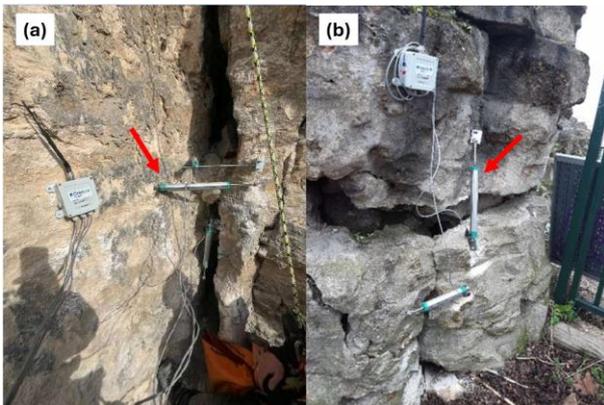

Figure 3. Example of electronic extensometers installed on a 30 cm vertical crack in the gypsum on the western side of the island (fissure 1, a), and on a 15 cm horizontal crack in masonry walls below the Temple of Sybil (fissure 2, b). The red arrow points at the extensometers analyzed in this work.

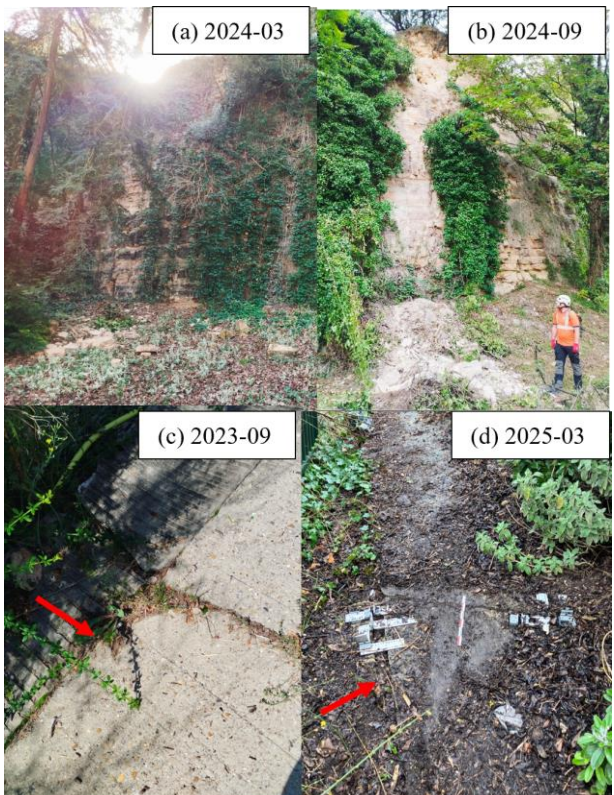

Figure 4. Observations during field visits. Top: examples of rockfalls. Bottom: Pathway photographed in September 2023 and March 2025, showing crack evolution (between pathway and gutter, red arrows) and manual gauges installed in 2024 (bottom right)

## 4 RESULTS

### 4.1 *Quarterly visits*

Quarterly visits include an overview of the whole park but focus on the island sector. 10 visits were carried out between March 2023 and June 2025. Regular rockfalls from the gypsum rock faces have been identified, most of them involving blocks of about 1 liter. However, at least 4 more important rockfalls (between 0.5 and 1m$^3$) also occurred, initiated from the top of the gypsum rock faces where roots development and associated meteoric water circulations favor the development of cracks (Figure 4a and b).

The movements of a shallow landslide on the island were also monitored (Figure 4c and d). Located in marls and probably associated with the ageing of historical geotechnical structures, its initiation date is not known. However, the BRGM witnessed its reactivation starting between September and December 2023, as evidenced by growing gaps between an inclined pathway and adjacent rainwater drainage channels.

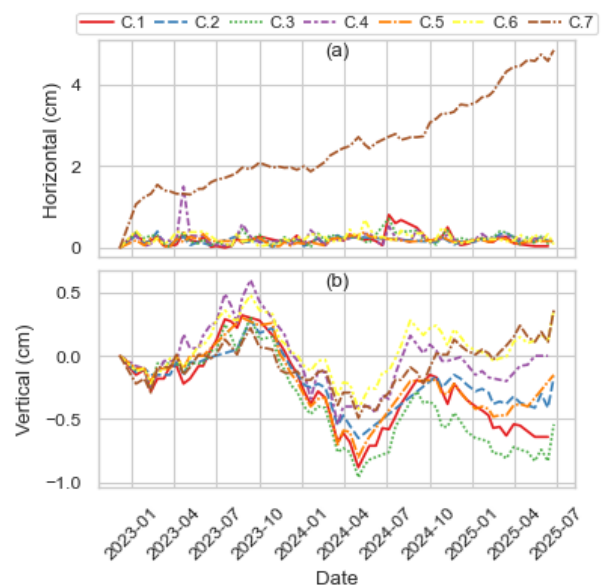

Figure 5. Horizontal (a) and vertical (b) displacements measured with tacheometric surveys since 2023 on 7 targets (C.1 to C.7) on the western side of the island (red ellipse in Figure 2).

### 4.2 *Topographic surveys*

The topographic surveys carried out since December 2022 aim at detecting (1) detachments of individual compartments already partly detached from the masonry walls and/or rock faces and (2) generalized movements at the scale of the rock faces/ masonry walls. The expected precision, as given by the surveyor, is of about 1 mm. In practice, displacements of up to 0.5 – 1cm were observed both horizontally and vertically, without being necessarily associated to clear trends. Only one target, located on purpose on a ~1 liter gypsum block partially detached from the rock face, showed a clear and homogenous horizontal movement away from the rock face (totalizing about 2.5 cm between January 2023 and June 2025, Figure 5a). This steady displacement accelerated in October 2024, from an average ~1 to 2 mm/month. There is, however, no sign of imminent rupture. Besides, no significant vertical displacement is measured.

No clear trend is observed on other targets (individually or as a group). Homogeneous displacements on nearby targets were sometimes observed during several weeks / months but were then followed by displacements in the opposite direction. The only homogeneous behavior observed on almost all targets was vertical displacements, with upward movement from April to October, and downwards movements from October to April (Figure 5b). The amplitude of these movements varies between the targets and the period. For instance, on the western side of the island, targets moved about 5 mm upwards in April – October 2023 and 7 mm in April – October 2024, and 1 cm downwards in October 2023 – April 2024 and only 2.5 mm in October 2024 – April 2025. In the same period, on the eastern side of the island, up to 7 mm of downward displacement was measured. Over the 60 targets and 2.5 years, a global trend of downward movement seems to emerge (on average 0.5 to 1 mm since January 2023).

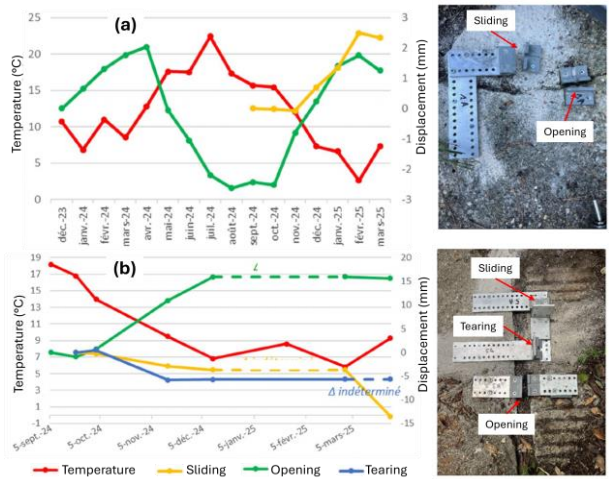

Figure 6. Manual gauges measurements on two cracks located on pathways on the island, along with ambient temperature measurements. Dashed lines on the bottom subfigure correspond to measurement gaps for which the displacements are not known. The bottom graph corresponds to the gauges in Figure 4d.

### 4.3 Manual gauges measurements

Manual measurements carried out on specific instrumented fissures provide more precise measurements of opening, sliding and tearing. Measurements are carried out on wedges fixed on both sides of the cracks with a digital fissure gauge, with an announced precision of 0.01 mm and a measurement range of 27 mm. Apart from a few seemingly aberrant measurements (i.e. significantly different of the mean of previous and following measurements), measurements are relatively stable which facilitates interpretation. The most significant displacements are measured on pathways, with opening/closing amplitudes of 3 to 4 mm (Figure 6a). Gauges installed on masonry walls display displacement of less than 1mm. These variations are associated to clear seasonal trends, with cracks opening during autumn/winter, and closing during spring/summer. The only place where clear, non-seasonal, displacement trends are observed is on the path going through the shallow landslide mentioned in section 4.1, with crack opening of more than 1.5 cm and exceeding the gauges measurement range (Figure 6b).

### 4.4 Extensometers measurements

The automatic, real-time measurements carried out with electronic extensometers aim at identifying displacement trends leading to rupture on major compartments identified as potentially unstable by the BRGM, with volumes estimated between 10 and 1,000 m$^3$. In combination with temperature probes placed at the surface and 50 cm deep in the rockface, they also allow to quantify the correlation between temperature variations and cracks opening / closing.

Since the installation of the extensometers in March 2024, there has been no evidence of opening/closing trends heralding short-term or medium-term collapse. However clear short-term and medium-term cycles can be identified, with cracks opening when temperature decreases (because of thermal contraction of rock/masonry compartments) and closing when temperature increases (as a result of thermal dilatation of compartments).

This is clearly visible in the short-term with day/night temperature variations for *fissure 1* (Figure 7b). Associated opening/closing cycles are all the more important as temperature differences are significant (Figure 8a). The correlation is less clear with temperature measured at depth (Figure 8b), suggesting these processes are related to surface mechanisms. The most important daily opening/closing amplitudes are mostly associated with sunny days (Figure 8c) with little rain (Figure 8c). Interestingly, these daily cycles are clearly visible when the cracks are closing in the medium term but are less visible when the crack is opening.

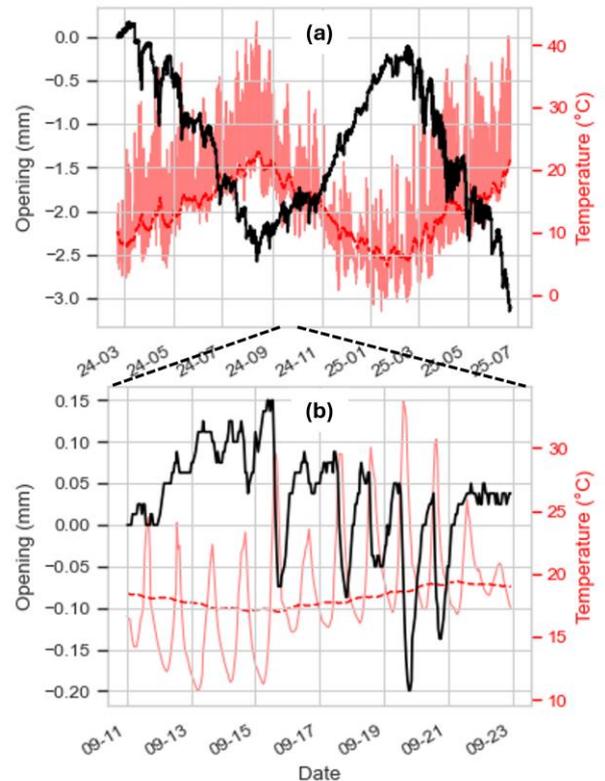

Figure 7. Automatic extensometer measurements at fissure 1 (Figure 3a). Black: opening time series. Light red curve: temperature of the extensometer digitizer. Dashed red curve: Temperature at 50cm depth. Top: Times series between March 2024 and June 2025 (date format is year-month). Bottom: time series between 9 and 23 September 2024 (date format is month-day).

The correlation between temperature and crack opening is also clear in the medium-term with cracks closing in spring/summer and opening during autumn/winter (Figure 7a). For *fissure 1*, a simple linear relation can be derived between the opening/closing of the crack, and the temperature at 50 cm depth within the rock face (Figure 9a). The crack closing/opening rate is ~0.2 mm/°C, but deviations around this trend lead to an irreversible behavior: the crack closed by about 0.5 mm between the summers of 2024 and 2025. As for daily cycles, significant precipitations seem to reduce the closing/opening velocity (Figure 9b), but no clear relation can be drawn.

Daily opening/closing cycles are also visible at *fissure 2* and the autumn/winter period is also associated with crack opening (Figure 10). However, contrary to *fissure 1*, the opening phase started in June 2024 and May 2025 while temperatures were still increasing. In turn, the fissure seems to follow a pluri-annual opening trend (about 2 mm between May 2024 and May 2025). Nonetheless, at least another full annual cycle is required to confirm this trend.

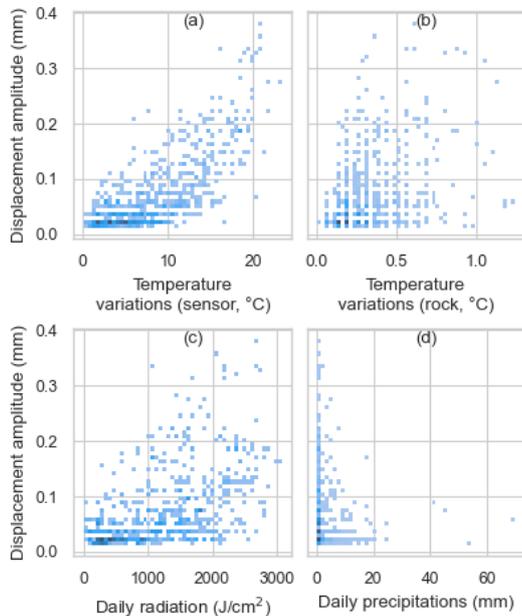

Figure 8. Normalized distribution of daily opening/closing amplitude at fissure 1 (Figure 3a) and daily temperature variations of the sensor (a), at 50 cm depth (b), daily radiation (c) and daily precipitations (d) at the Luxembourg park meteorological station (Météo-France, 2025).

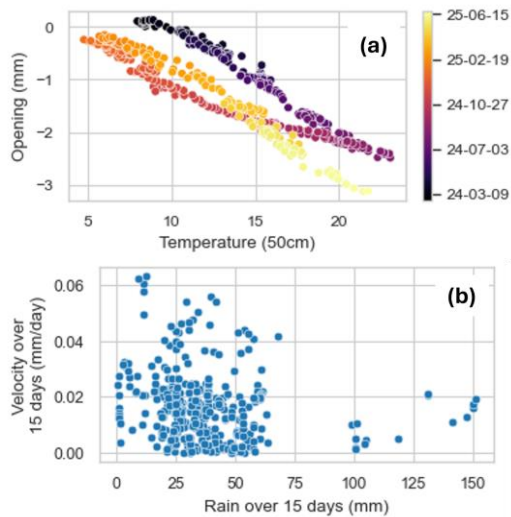

Figure 9. Analysis of displacements at fissure 1 between March 2024 and June 2025 (Figure 3a). (a) Mean daily opening (y-axis) vs mean daily temperature (y-axis) measured at 50 cm depth (date format is year-month-day in the colorbar). (b) Velocity computed over a 15 days time window (y-axis) vs precipitations over the previous 15 days at the Luxembourg park meteorological station (Météo-France, 2025).

## 5 DISCUSSION

The complexity of the geotechnical context and expected hazards, as well as the size of the site, requires complementary methods to monitor the sector of the island in the Buttes Chaumont Park.

The quaterly field visits primarily rely on visual observations to identify key evolutions in the different sectors as an exhaustive program of quantified measurements cannot be implemented in the entire monitored sector. The repetition of the visits over a long period (more than 2 years at the time of writing) contributes to (i) construct a catalog of located and dated disorders occurring in the park (rockfalls, evolution of cracks, …) and (ii) developing an expert-based knowledge and understanding of the Buttes Chaumont site, which is necessary to interpret both identified evolutions and quantified measurements, and in turn suggest adequate mitigation measures when necessary. In particular, the BRGM:

- recommended to establish a 10 m security perimeter above the gypsum rock face, following the rockfalls that occurred between December 2023 and March 2024;
- identified cracks to be instrumented with manual gauges or electronic extensometers, at the beginning of the instrumentation and later on to complement it following observed evolutions;
- noted some structural disorders in other ageing structures (e.g. retaining walls) in the park that do not require immediate repairs but that are worth characterizing by future, more detailed, geotechnical diagnostics.

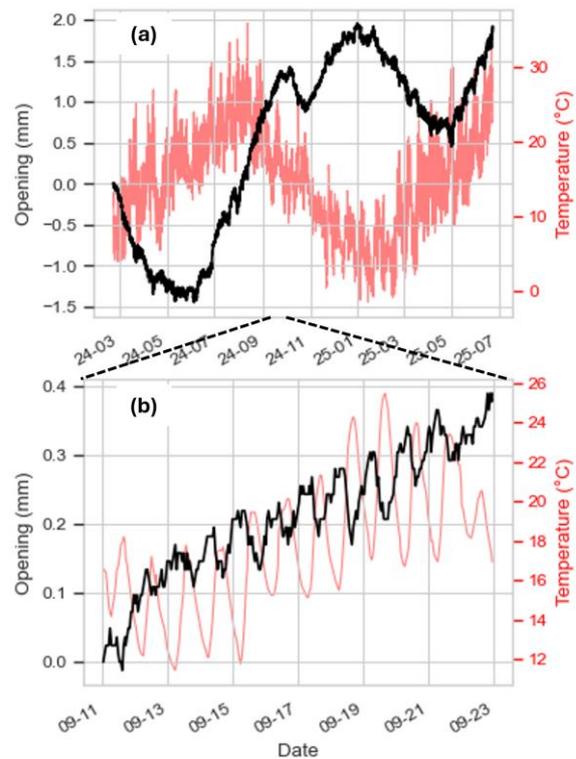

Figure 10. Automatic extensometer measurements at fissure 2 (Figure 3b). Displacement time series are given in black. The light red curve is the temperature recorded by the extensometer digitizer. Top: Times series between March 2024 and June 2025 (date format is year-month). Bottom: time series between 9 and 23 September 2024 (date format is month-day).

More generally, the observations carried out during field visits largely contributed to raising awareness among stakeholders about the geotechnical hazards in the island sector, illustrating the importance of protection measures taken from 2021 onwards (e.g. closure of the island to the public and preliminary geotechnical investigation), as well as the need for an ambitious restoration program.

The field visits are complemented by quantified displacement measurements to monitor more finely the evolution of rock faces and masonry walls. The relative low precision of tacheometer measurements (about 2 – 3 mm) does not allow to monitor closely these evolutions. Displacements trends vary from one month to another with no easy interpretation and possible measurement bias. For instance, we identified two targets with east/west annual displacements cycles of 1 cm amplitude. The interpretation of these cycles is difficult as they are not located on the same structure, and targets located close to them do not display a similar behavior. Nevertheless, significant movements (above 1 cm) of compartments comprising several targets should be detectable, which would then allow to take appropriate mitigation measures. The annual upward/downward movement cycles identified in this work must be confirmed by future acquisitions, as for the average 0.5 – 1 mm downward movement observed over the past 2.5 years. The latter is however of the same order of magnitude as INSAR measurements (Copernicus, 2025) of about 1 mm / year between 2019 and 2023 around the Buttes Chaumont Park.

Annual displacement cycles are also evidenced by manual gauges measurement and electronic extensometers. Manual gauges measurements helped quantify displacements on a shallow landslide on the island, thus objectifying visual observations. The displacements were, however, too important to be tracked by the gauges that had to be re-calibrated, interrupting the continuity of measurements. At the time of writing, the installation of new instrumentation is planned, including manual gauges with longer strokes and electronic extensometers. The installation of borehole inclinometers was also considered but is not feasible as, for safety reasons, drilling machines cannot be simply brought on site.

The electronic extensometers were primarily installed to monitor short-term evolution of potentially unstable rock/masonry compartments. An alarm system has been implemented based on absolute displacement values and velocity measured on an hourly, weekly and monthly basis. So far, no alarming signs of rupture have been identified. The seasonal / daily cycles of cracks opening/closing are, however, clear indications of the control of temperature on the dynamics of fissure.

Such observations are of prior importance to distinguish between reversible (associated with rock dilatation/contraction) and irreversible (associated with detachment) movement of the compartments. Interestingly, rain appears to have a limited influence on the dynamics of cracks, as significant precipitations are associated to small daily opening/closing amplitudes and, in a lesser extent, small medium-term open or closing velocities. This may be related to the fact that rainy days are cloudy and thus associated with limited temperature variations, which primarily controls daily opening/closing cycles. In the medium term, rain may also contribute to homogenizing the temperature of the gypsum massif, thus limiting differential dilation/contraction and fracture opening/closing. This hypothesis should however be further investigated by a systematic analysis of correlations between crack opening, temperature and precipitation, for the different instrumented fissures.

## 6 CONCLUSIONS

The Buttes Chaumont Park is a unique geotechnical complex resulting from the development of a public park in a former gypsum quarry. It was designed to mimic the romantic vision of nature of the second half of the 19th century, and as such works included the construction of an artificial lake and cavern, and of artificial rock faces. The ageing of the structures (retaining walls, cladding, …) combined with the unstable nature of gypsum results in multiple geotechnical and geological hazards, with in particular rockfalls (from both gypsum and artificial structures) and landslides.

In order to characterize these hazards, monitor their evolution and support the Paris city council in their ambitious restoration project, the BRGM has designed and supervised a monitoring scheme for the island sector of the park. This monitoring is based on complementary approaches including regular field visits with visual observations, and quantitative displacement measurements with tacheometric surveys, manual gauges measurements and automatic electronic extensometers measurements. Over the past 2.5 years, this monitoring scheme has significantly improved the understanding of gravitational processes on the rock and masonry faces. Significant rockfalls (up to 1 m$^3$) have occurred over the past two years, and a shallow landslide is active on the island (with no sign of imminent rupture) which supports the 2021 decision to restrict access of some areas. Fortunately, there has been so far no evidence of imminent rupture of rocky/masonry compartments identified as potentially unstable.

Through the monitoring scheme the BRGM has in particular:

- Constructed a database of located and dated events (rockfalls, fissure opening, landslide displacements, disorders in geotechnical structures) that complements previous historical studies and highlights the need for a restoration program;
- Helped the Paris city council interpret observations and take appropriate mitigation measures;
- Carried out a preliminary, quantitative analysis of correlations between meteorological parameters and opening/closing fissures in gypsum faces and masonry walls. This helps understanding reversible processes at stake in the evolution of fissures to, in turn, better characterize irreversible processes that could lead to collapses.

These findings can be complemented by future, more detailed analyses of displacement time series. The hazard assessment and associated recommendations will have to be constantly re-evaluated with new data and observations to come. At last, but not least, this unique, several years long, observation chronicle will play a major role in restoration works to come. Indeed, it provides a baseline that will guide technical choices for future safety works and allow to assess their effectiveness.

## 7 ACKNOWLEDGEMENTS

We thank the DEVE (Direction des Espaces Verts) service of Paris city council for granting access to data related to this study and for their cooperation for this publication. Tacheometer measurements presented in this work are carried out by the PROGEXIAL company. The Ginger company is responsible for manual gauges measurements and for the extensometers monitoring (installation, maintenance and data transfer).